\documentstyle[aps,preprint]{revtex}

\begin{document}
\draft
\title{Bubbles created from vacuum fluctuation\thanks{%
Supported by the National Natural Science Foundation of China under No.
19473005, and the Education Committee of Hunan Province.}}
\author{Liao Liu}
\address{Department of Physics, Beijing Normal University, Beijing 100875, P. R.\\
China\thanks{%
Email address: liaoliu@class1.bao.ac.cn or fhedoc@263.net}}
\author{Feng He}
\address{Department of Physics, Beijing Normal University, Beijing 100875, P. R. China%
\\
and\\
Department of Physics, Xiang Tan Normal College, Xiang Tan 411201, P. R.\\
China}
\maketitle

\begin{abstract}
We show that the bubbles $S^2\times S^2$can be created from vacuum
fluctuation in certain De Sitter universe, so the space-time foam-like
structure might really be constructed from bubbles of $S^2\times S^2$ in the
very early inflating phase of our universe. But whether such foam-like
structure persisted during the later evolution of the universe is a problem
unsolved now.
\end{abstract}


J.A.Wheeler was the first one who pointed out that the space-time might have
a foam-like structure around the Plank scale\cite{1},though it is simply
connected and smooth at large scale. But what is the constituents of the
space-time foam is a problem long unsolved. Hawking himself oscillated
between the wormhole pictured and bubble pictured foam-like structure\cite{2}%
.However according to our point of view, the essential point is whether the
bubble or wormhole could be really solutions of the semi-classical Einstein
gravitational equation

\begin{equation}
G_\mu ^\nu =-\frac{8\pi G}{c^4}\langle 0\mid T_\mu ^\nu \mid 0\rangle _{ren},
\end{equation}
where $G_\mu ^\nu =R_\mu ^\nu -\frac 12R\delta _\mu ^\nu $ is the Einstein
tensor of the bubbles or wormhole, $\langle 0\mid T_\mu ^\nu \mid 0\rangle
_{ren}$ is however the renormalized matter stress-energy tensor of vacuum
fluctuation in certain given classical background space-time, $G$ is
gravitational constant, $c$ is velocity of light. The signature in our paper
is -2.

Early in 1993, one of the authors (L.Liu) had found a transient Lorentzian
wormhole solution\cite{3}

\begin{equation}
ds^2=d(ct)^2-l_p^2\cosh ^2(ct/l_p)d\Omega _3^2,
\end{equation}
where $l_p$ is the Planck length. From (1), this Lorentzian mini-wormhole
creates at certain early time and annihilates later under vacuum fluctuation
in a closed inflating de Sitter background space-time of metric

\begin{equation}
ds^2=d(ct)^2-\alpha ^2e^{ct/\alpha }d\Omega _3^2,
\end{equation}
where $\alpha =\sqrt{3/\Lambda }$, $\Lambda $ is the cosmological constant
of the background de Sitter universe. Now the problem is whether the
Euclidean bubble $S^2\times S^2$ can also be a solution of the
semi-classical Euclidean Einstein field equation (1)? As is known, the
Euclidean metric of $S^2\times S^2$ is the Nariai instanton of metric\cite{4}

\begin{equation}
ds^2=-\lambda ^{-1}(\sin ^2\chi d\stackrel{\_}{\psi }^2+d\chi ^2+d\Omega
_2^2),
\end{equation}
where $\stackrel{\_}{\psi }$ is the imaginary time and $\sqrt{1/\lambda }$
is the radius of the 2-sphere $S^2$. $\lambda $ in our paper is nothing to
do with the so-called cosmological constant, though it is so in the
degenerated De Sitter-Schwarzschild spacetime historically. The Lorentzian
Nariai metric is \cite{4}

\begin{equation}
ds^2=\lambda ^{-1}\sin ^2\chi d\psi ^2-\lambda ^{-1}d\chi ^2-\lambda
^{-1}d\theta ^2-\lambda ^{-1}\sin ^2\theta d\varphi ^2),
\end{equation}
where $\psi =-i\stackrel{\_}{\psi }$ is a real time variable.

From the Lorentzian Nariai metric (5) and (1), the renormalized vacuum
matter stress-energy tensor can be obtained straightforward as (Appendix)

\begin{equation}
\langle 0\mid T_\mu ^\nu \mid 0\rangle _{ren}=-\frac{c^4}{8\pi G}G_\mu ^\nu =%
\frac{c^4}{8\pi G}\lambda \delta _\mu ^\nu .
\end{equation}

However expression (6) is just the well-known vacuum matter stress-energy
tensor in one-loop approximation for scalar field\cite{5}

\begin{equation}
\langle 0\mid T_\mu ^\nu \mid 0\rangle _{ren}=\frac{\hbar c}{960\pi ^2\alpha
^4}\delta _\mu ^\nu =\frac{c^4\Lambda ^2l_p^2}{8640\pi ^2G}\delta _\mu ^\nu
,\quad (\alpha \equiv \sqrt{\frac 3\Lambda },\ \Lambda \text{ is the
cosmological constant}).
\end{equation}
of the inflating flat de Sitter universe

\begin{equation}
ds^2=d(ct)^2-e^{\frac{2ct}\alpha }dx^idx_i
\end{equation}
or the inflating closed de Sitter universe

\begin{equation}
ds^2=d(ct)^2-\alpha ^2\cosh ^2(ct/\alpha )d\Omega _3^2
\end{equation}

If we put

\begin{equation}
\lambda =\frac{l_p^2\Lambda ^2}{1080\pi },
\end{equation}
if the bubbles $S^2\times S^2$ are the result of the vacuum fluctuation of
certain background de Sitter space-time, then a reasonable demand of (10) is 
$\sqrt{\lambda ^{-1}}\ll \sqrt{3\Lambda ^{-1}},\ or\quad \lambda \gg \Lambda
/3$, that is $\Lambda \gg \frac{1080\pi l_p^{-2}}3\sim 10^{69}$, here we
would like to point out again though $\Lambda $ is the cosmological constant
of the background De Sitter spacetime, $\lambda $ is not, $\lambda $ relates
only with the radius of the two sphere $S^2$.

In concluding, we show unambiguously that the bubbles of topology $S^2\times
S^2$ can be really created from vacuum fluctuation in the background de
Sitter universe of $k=0,1$.

However important comments should be given now, i.e., First, in our
understanding, it seems the instanton $S^2\times S^2$ is just certain kind
of compact topological object with $\chi =4$ and index $\tau =0$, which have
no connection with the creation of black hole pairs. In fact, we agree with
Bousso and Hawking\cite{4}: ``Strictly speaking, it does not even contain a
black hole, but rather two acceleration horizons.''Second, It seems either
the wormhole pictured or the bubble pictured space-time foam-like structure
can only be created from the vacuum fluctuation in the inflationary era of
our universe. This remark had already been pointed out by Buosso and Hawking%
\cite{4} and Hawking\cite{2}, but whether such spacetime foam-like strucure
stably exist during the later evolution of our universe is an interesting
problem unsolved. So it seems that the stability of the above mentioned
foam-like structure should be studied in order to confirm that
Wheeler--Hawking's conjecture is true or not.

\acknowledgments 

This work is supported by the National Natural Science Foundation of China
under Grant No.19473005 and the Education Committee of Hunan Province.

\appendix 

\section{Appendix}

The Lorentzian Nariai metric of $S^2\times S^2$ is

\[
ds^2=\lambda ^{-1}\sin ^2\chi d\psi ^2-\lambda ^{-1}d\chi ^2-\lambda
^{-1}d\theta ^2-\lambda ^{-1}\sin ^2\theta d\varphi ^2 
\]
namely $g_{00}=\lambda ^{-1}\sin ^2\chi ,$ $g_{11}=-\lambda ^{-1},$ $%
g_{22}=-\lambda ^{-1},g_{33}=-\lambda ^{-1}\sin ^2\theta ,$ $g_{\alpha \beta
}=0(\alpha ,\beta =0,1,2,3;\alpha \neq \beta )$. The determinant of the
metric is $g=-\lambda ^{-4}\sin ^2\chi \sin ^2\theta $.

From $g^{\mu \lambda }=\Delta ^{\mu \lambda }/g$, we can compute
contravariant components of the metric as follows: $g^{00}=\lambda \sin
^{-2}\chi ,$ $g^{11}=-\lambda ,$ $g^{22}=-\lambda ,g^{33}=-\lambda \sin
^{-2}\theta ,$ $g^{\alpha \beta }=0(\alpha ,\beta =0,1,2,3;\alpha \neq \beta
)$. From $\Gamma _{\mu \nu }^\alpha =\frac 12g^{\alpha \lambda }(g_{\mu
\lambda ,\nu }+g_{\nu \lambda ,\mu }-g_{\mu \nu ,\lambda }),$ we get the
non-zero components of Christoffel which are $\Gamma _{01}^0=\Gamma
_{10}^0=ctg\chi ,$ $\Gamma _{00}^1=\sin \chi \cos \chi ,\Gamma _{33}^2=-\sin
\theta \cos \theta ,$ $\Gamma _{23}^3=\Gamma _{32}^3=ctg\theta $.

Put the values of the above Christoffel into the following formula of Ricci
tensor $R_{\mu \nu }=R_{\mu \lambda \nu }^\lambda =\Gamma _{\mu \lambda ,\nu
}^\lambda -\Gamma _{\mu \nu ,\lambda }^\lambda +\Gamma _{\sigma \nu
}^\lambda \Gamma _{\mu \lambda }^\sigma -\Gamma _{\sigma \lambda }^\lambda
\Gamma _{\mu \nu }^\sigma ,$ we can get the non-zero components of Riccci
tensor as follows $R_{00}=\sin ^2\chi ,$ $R_{11}=-1,$ $R_{22}=-1,$ $%
R_{33}=-\sin ^2\theta $. We can also compute Ricci scalar as follows $%
R=g^{\mu \nu }R_{\mu \nu }=4\lambda $. Through the formulas $G_{\mu \nu
}=R_{\mu \nu }-\frac 12g_{\mu \nu }R$, we get the non -zero components of
Einstein tensor which are $G_{00}=-\sin ^2\chi ,$ $G_{11}=1,$ $%
G_{22}=1,G_{33}=\sin ^2\theta $. From the formulas $G_\mu ^\nu =g^{\nu \rho
}G_{\rho \mu }$ ,we get $G_\mu ^\nu =-\lambda \delta _\mu ^\nu $.

\end{document}